# Superfluorescence in CdS/CdSe/CdS Spherical Quantum Wells Modulated by Excitation Geometry

Chunzheng Bai[1], Qihao Sun[1], Yiying Zhu[1], Ruilin Xu[2]*, Bin Gu[1], Jinsong Qi[3], Fengrui Hu[3], Xiaoyong Wang[3], Jiayu Zhang[1]*

[1] *School of Electronic Science and Engineering, Southeast University, Nanjing 210096, China*

[2] *School of Information Science and Technology, Nantong University, Nantong 226019, China*

[3] *Nanjing University, Nanjing 210093, China*

* *xrl@ntu.edu.cn*, *jyzhang@seu.edu.cn*

**Abstract**

Superfluorescence (SF) originates from the spontaneous buildup of macroscopic coherence among initially incoherent emitters. Its formation is highly sensitive to dephasing and collective coupling in the system. Here, we observe room-temperature SF in CdS/CdSe/CdS spherical quantum wells. With increasing excitation fluence, the peak emission intensity shows a nearly quadratic increase, while the emission delay and pulse width decrease. Burnham–Chiao ringing is also observed, revealing the characteristic collective radiation dynamics of SF. Further experiments with stripe excitation show that shortening the excitation length $L$ changes the dominant fast emission from amplified spontaneous emission to SF. More importantly, the threshold for SF under stripe excitation is only about 1/30 of that under spot excitation. This large reduction indicates that the propagating radiation field provided by spatially extended excitation favors the buildup of cooperative coherence. These findings provide direct experimental evidence for understanding ultrafast many-body coherence dynamics and offer a route to actively control collective emission at room temperature.

**Introduction**

Superfluorescence (SF) is a cooperative emission process in which initially incoherent excited emitters spontaneously develop macroscopic coherence and radiate collectively.[1–4] Because this coherence builds up from quantum fluctuations, SF shows a clear buildup delay. Its emission lifetime and delay both decrease markedly with increasing excitation density, while the peak intensity increases nearly quadratically with the number of cooperative emitters. SF is also often accompanied by characteristic Burnham–Chiao ringing.[5–9] However, macroscopic polarization must be established within the dephasing time $T_2$ ($\tau_{SF} < T_2$), making SF highly sensitive to optical dephasing. Therefore, early observations of SF required demanding experimental conditions. For example, cooperative emission in InGaAs/GaAs multiple quantum wells required temperatures of ~ 5 K and magnetic fields above 10 T to suppress phonon scattering and carrier dephasing.[10] Even in highly ordered $CsPbBr_3$ nanocrystal superlattices, SF was initially observed only at about 6 K.[6] In recent years, SF has been extended to room temperature in lead halide perovskites. This progress has been supported by their distinct exciton–lattice coupling and high-density many-body states.[8,11] These studies have improved our understanding of cooperative coherence at room temperature and have provided a basis for further exploring the conditions that govern SF in semiconductors.

On this basis, an important question in semiconductor SF research is to further clarify the conditions for the establishment of cooperative coherence in the presence of competing radiative channels and to control the resulting collective emission state. Previous studies have shown that increased dephasing at higher temperatures can drive a continuous transition from SF to amplified spontaneous emission (ASE). In semiconductor quantum wells, increasing the magnetic field and excitation density can instead shift the emission from an ASE-dominated regime to an SF-dominated regime.[12] Recently, Kobiyama *et al.* showed in giant $CsPbBr_3$ nanocrystal films that temperature and exciton density can regulate the transitions among SF, ASE, and photoluminescence (PL). They also found that changing the excitation-stripe length strongly affects the emission state and transient dynamics under high excitation.[13] These studies show that the final radiative state depends on the balance among the growth of cooperative coherence, dephasing, optical-field effects, and other competing radiative processes. Notably, the spatial geometry of the excitation region acts in a way different from temperature and carrier density. It not only defines the group of emitters involved in the emission, but also directly changes photon propagation within the excited region. Therefore, the spatial extent of excitation affects the buildup of propagation gain and may also modify the formation of cooperative coherence. It is thus an important factor for understanding different fast radiative states and the transitions between them.

Here, we use CdS/CdSe/CdS spherical quantum wells (SQWs) as the model system and vary the length of the stripe excitation region to examine how the fast emission dynamics evolve with the spatial extent of excitation at room temperature. The results show that, as the excitation region becomes shorter, the dominant fast emission changes from ASE to SF. Further comparison with spot excitation shows that a finite-length stripe excitation can greatly reduce the threshold for SF. These results indicate that the excitation length influences both the dominant fast emission process and the buildup of cooperative coherence in SF. They also provide experimental evidence for how excitation geometry affects the formation of cooperative emission.

**Results and Discussion**

CdS/CdSe/CdS SQWs were synthesized through layer-by-layer shell growth and then deposited on quartz substrates as thin films. To investigate the collective cooperative emission dynamics and macroscopic quantum coherence in the films, excitation-dependent time-resolved PL (TRPL) spectra were measured using a streak camera under 400 nm fs excitation. Fig. 1a–c show the TRPL spectra of the SQWs at different excitation fluences. At low excitation fluence, the emission shows typical spontaneous emission (SE) (Fig. 1a). When the excitation fluence is increased to 7.06 and $42.36\times10^{3}$ μJ $cm^{-2}$, the emission develops into a short, intense transient pulse. With further increasing excitation fluence, the pulse width of the main pulse decreases markedly (Fig. 1b,c). Meanwhile, the main pulse is followed by distinct delayed oscillations, showing transient dynamics clearly different from those of low-excitation SE.

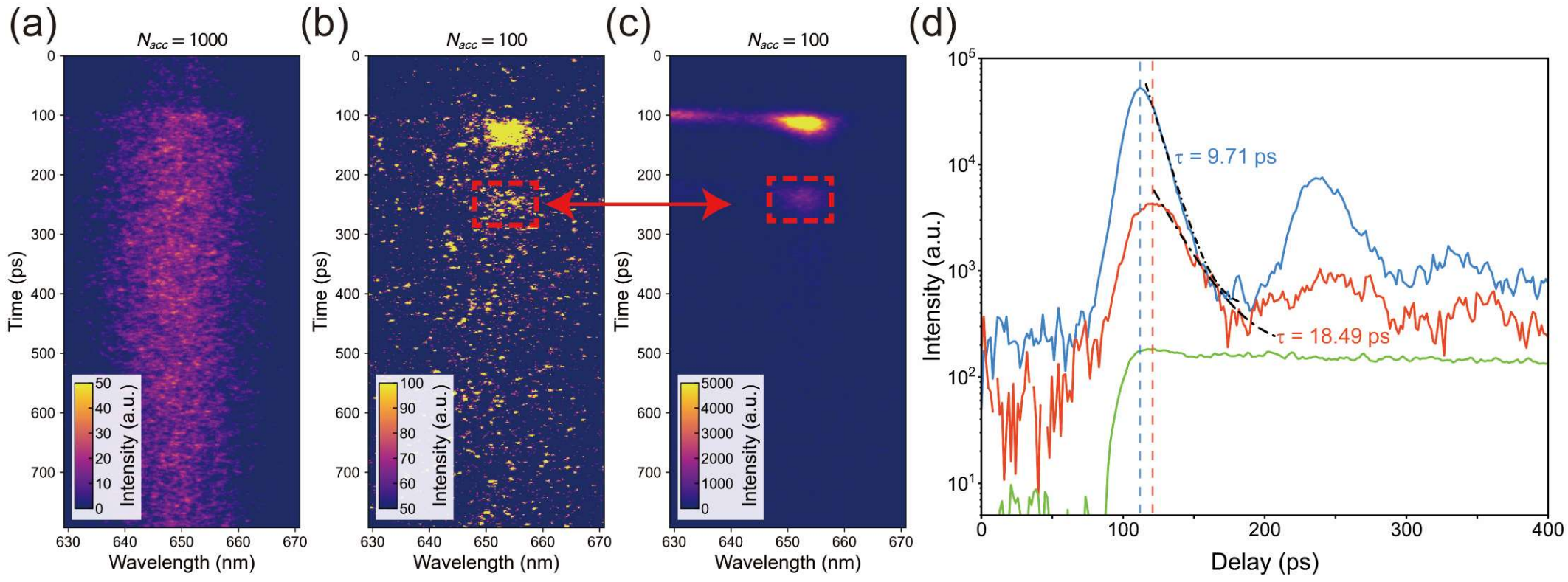

**Fig. 1 Excitation-dependent emission dynamics of CdS/CdSe/CdS spherical quantum-wells**

**(SQWs) films. a–c,** Time-resolved PL (TRPL) spectra at increasing excitation fluences. **d,** Corresponding spectrally integrated TRPL traces. $N_{\mathrm{acc}}$ denotes the number of accumulations. Red dashed boxes in **b,c** highlight the delayed oscillations following the main pulse.

This evolution is more clearly seen in the spectrally integrated TRPL traces in Fig. 1d. The peak intensity rises by nearly three orders of magnitude from the low-excitation SE level. Under high excitation, the traces show a dominant main pulse followed by damped oscillatory peaks, while no such oscillations are observed at low excitation. This damped temporal response is consistent with the characteristic Burnham–Chiao ringing of cooperative emission and indicates that coherent radiation dynamics persist after the main pulse.[6,7,14] In addition to the ringing, a clear emission delay is also observed. The two dashed lines in Fig. 1d mark the peaks at the higher excitation fluences. Both occur later than the low-excitation SE response, indicating a finite buildup process before the strong transient emission. As the excitation fluence increases, the main peak shifts earlier by about 9 ps. This shorter delay reflects a reduced buildup time at higher excitation density, indicating a faster buildup of cooperative coherence. At the same time, the decay also shows a clear dependence on excitation fluence. The initial decay after the main peak becomes faster at higher excitation, with the 1/e decay time decreasing from about 18 to 10 ps. This reduction indicates an acceleration of the radiative dynamics as the excitation increases. Together with the finite buildup delay and Burnham–Chiao ringing, it further supports the cooperative nature of the emission.[15,16]

To further quantify the cooperative emission dynamics, we analyzed the pulse profiles at different excitation fluences. TRPL dynamics can generally be described by exponential models.[17,18] For cooperative emission with oscillatory features, damped-oscillation models or Maxwell–Bloch equations can be used to analyze the coherent dynamics.[7,19] For an ideal SF pulse, the SF model predicts a characteristic $\mathrm{sech}^2$ temporal profile. This profile has been widely used to quantify the pulse width, peak intensity, and delay time of SF.[2,3,20] We therefore fitted the transient emission pulses with a $\mathrm{sech}^2$ function.

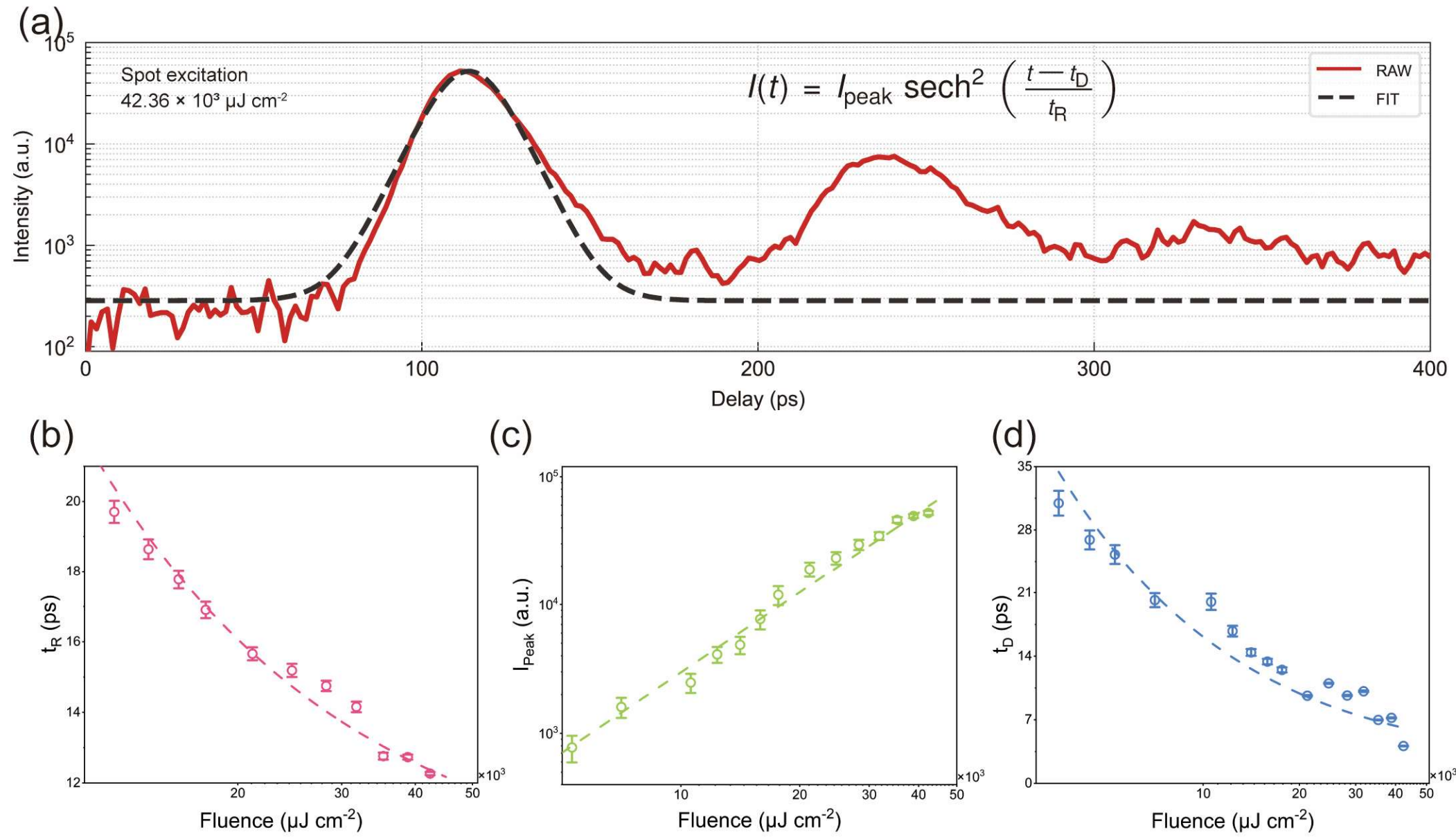

**Fig. 2 Excitation-dependent SF dynamics. a,** Representative spectrally integrated TRPL trace and corresponding fit. **b–d,** Characteristic radiation time $\tau$, peak intensity $I_{\text{peak}}$, and buildup time $I_{\text{peak}}$ as a function of excitation fluence, respectively. Dashed lines are fits to the corresponding excitation-fluence dependences.

Fig. 2a shows a representative temporal trace and its fit at an excitation fluence of $42\times10^3$ µJ cm$^{-2}$. The characteristic emission time $\tau$, peak intensity $I_{\text{peak}}$, and buildup time $t_{\text{b}}$ extracted using the same procedure are plotted in Fig. 2b–d. As the excitation fluence increases, the integrated intensity remains nearly linear, following $I_{\text{int}} \propto F^{1.02}$. This behavior indicates no obvious excitation-dependent nonradiative loss. Meanwhile, $\tau$ decreases from about 20 to 12 ps, showing the near $1/N$ scaling expected for SF cooperative emission (Fig. 2b). The peak intensity shows a strong superlinear increase. A log–log fit gives $I_{\text{peak}} \propto F^{2.06}$, corresponding to a nearly quadratic fluence dependence (Fig. 2c). An even stronger excitation dependence is observed in the coherence buildup. As the excitation fluence increases, $t_{\text{b}}$ rapidly decreases from about 31 to 6 ps and follows the $\ln N\,/N$ scaling expected for SF buildup dynamics (Fig. 2d).

The emission time, peak intensity, and buildup time all exhibit a correlated excitation dependence. In a cooperative system of $N$ emitters, the radiative time shortens as $N$ increases, while the collective polarization drives the peak intensity toward an $N^2$ scaling. Crucially, SF initiates from an ensemble of emitters with initially uncorrelated phases; the formation of an intense burst therefore requires the spontaneous buildup of macroscopic polarization, resulting in a finite buildup time. As the number of cooperative emitters increases, the coherence buildup accelerates and $t_b$ decreases accordingly. Together with the delayed pulses and Burnham–Chiao ringing observed in Fig. 1, these findings provide further evidence for SF in the SQWs. Because SF relies on the spontaneous buildup of macroscopic polarization among initially incoherent emitters, the growth of cooperative polarization must therefore be faster than optical dephasing.[5,21] For two-dimensional (2D) electron–hole system in semiconductor quantum wells, this competition can be described by the cooperative frequency $\omega_c$, and the condition for SF can be written as[22,23]

$$\omega_c = \sqrt{\frac{8\pi^2 d^2 n \Gamma c}{\hbar \tilde{n}^2 \lambda L_{\mathrm{QW}}}} \gtrsim \frac{2}{T_2} \tag{1}$$

Here, $d$ is the transition dipole moment, $n$ is the 2D carrier density, $\Gamma$ is the mode-overlap factor between the optical field and the active region, and $L_{\mathrm{QW}}$ is the total quantum-well thickness. The expression shows that a larger transition dipole moment enhances the cooperative coupling strength, allowing macroscopic polarization to build up before dephasing occurs.

In SQWs, carriers in the CdSe quantum well experience radial confinement, while their wavefunctions can extend along the spherical well layer, forming quasi-2D carrier states that are confined radially but delocalized over the well surface.[24–26] In quantum-well systems, such extended exciton states favor a large oscillator strength.[27–29] Atomistic calculations for CdS/CdSe/CdS quantum-well nanoshells have also shown a strong enhancement of the oscillator strength. Because the oscillator strength follows $f \propto \Delta E|d|^2$, a larger oscillator strength corresponds to a larger effective transition dipole moment when the transition energies are similar. [30,31] On the other hand, the relatively large confinement volume of SQWs can suppress Auger recombination and maintain long-lived biexciton and multiexciton populations.[32,33] These results indicate that fast nonradiative loss under high excitation is suppressed, helping to maintain a high effective emitter density and thereby increase the cooperative growth rate.

Subsequently, by systematically varying the pump stripe length $L$, we further investigated the influence of the excitation region size on the transient emission. TRPL measurements reveal that the radiative dynamics under high excitation change significantly with $L$. Representative results for

$L$ = 3.0 and 2.1 mm are shown in Fig. 3.

Owing to the effective suppression of rapid non-radiative losses from highly excited states in SQWs, the system exhibits substantial optical gain. For $L$ = 3.0 mm, increasing the excitation fluence from 30 to 74 μJ $cm^{-2}$ leads to a rapid increase in emission intensity together with clear spectral narrowing (Fig. 3a–c). The corresponding spectrally integrated TRPL traces show that the emission peak remains at nearly the same delay over the fluence range (Fig. 3d). Consistently, the pump-dependent spectra show that the integrated PL intensity undergoes a superlinear increase above about 40 μJ $cm^{-2}$. These dynamical and intensity features indicate that ASE dominates the emission under long-stripe excitation.[26,34]

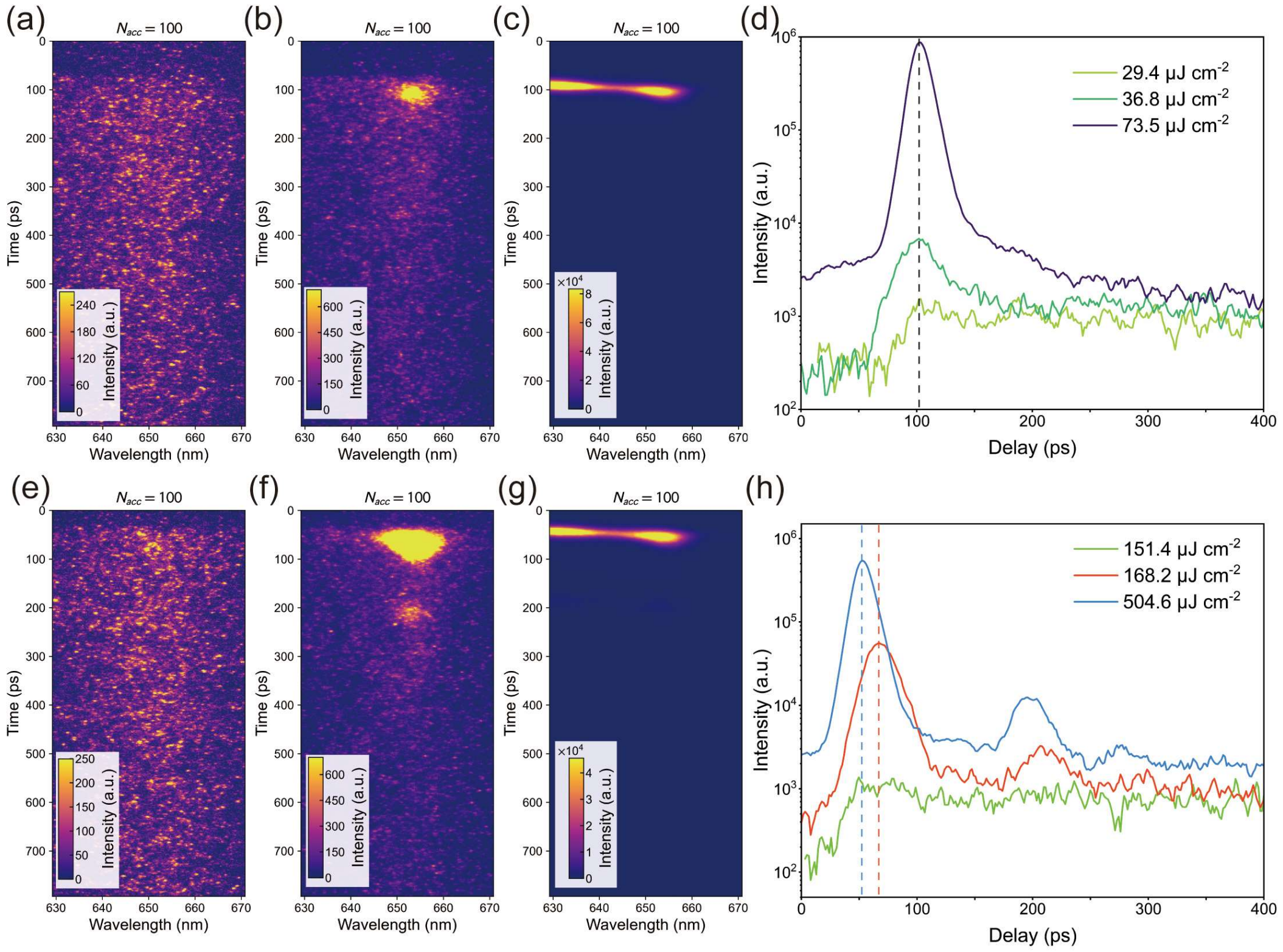

**Fig. 3 Excitation-length-dependent emission dynamics. a–c,** TRPL spectra at increasing excitation fluences for $L$ = 3.0 mm. **d,** Corresponding spectrally integrated TRPL traces. **e–g,** TRPL spectra at increasing excitation fluences for $L$ = 2.1 mm. **h,** Corresponding spectrally integrated TRPL traces.

When the excitation stripe is shortened to $L$ = 2.1 mm, the time-resolved emission shows distinctly different dynamics (Fig. 3e–g). As the excitation fluence increases, the emission narrows spectrally and resolvable ringing appears in the tail of the transient emission. The corresponding spectrally integrated TRPL traces further reveal a clear difference from the $L$ = 3.0 mm (Fig. 3h). Compared with the low-excitation PL response, the intense emission at 168 and 504 μJ $cm^{-2}$ both show a clear emission delay, indicating a finite buildup of macroscopic coherence before the intense emission develops. With further increasing excitation fluence, this delay decreases by about 15 ps, consistent with the excitation dependence of the SF buildup time discussed above. Meanwhile, at $L$ = 2.1 mm, the integrated PL intensity shows a nearly linear dependence on excitation fluence, without the strong superlinear increase observed at $L$ = 3.0 mm. The concurrent manifestation of the emission delay, a shorter buildup time at higher excitation, and Burnham–Chiao ringing indicates that the dominant fast emission process changes from ASE to SF as the excitation stripe is shortened from 3.0 to 2.1 mm.

This transition originates from the different responses of ASE and SF to the excitation stripe length. ASE relies on photon propagation and stimulated amplification along the excited region, and its intensity is therefore strongly affected by the effective gain length $L$.[35,36] When $L$ is shortened from 3.0 to 2.1 mm, the available gain path is reduced and becomes insufficient to sustain strong ASE. The system, however, still satisfies the conditions for cooperative emission in terms of emitter number, coupling strength, and dephasing. As a result, the dominant fast emission process changes from ASE to SF. Remarkably, the SF threshold at $L$ = 2.1 mm is only about 1/30 of that under spot excitation. This large difference indicates that a finite-length stripe excitation affects not only the gain length for ASE, but also the buildup of cooperative coherence. Under stripe excitation, the SE field can propagate along the pumped region and continuously interact with spatially distributed emitters. When $L$ becomes too short to sustain ASE, this propagating radiation field can still remain. Such a subthreshold propagating field may facilitate the buildup of macroscopic polarization[37], thereby greatly reducing the SF threshold in the finite-stripe geometry.

**Conclusion**

In summary, room-temperature SF is observed in CdS/CdSe/CdS SQWs films. Further measurements show that the spatial extent of the excitation region plays an important role in the competition between stimulated amplification and cooperative emission. As the stripe excitation length is reduced, the dominant emission process changes to SF. More importantly, the SF threshold under finite-length stripe excitation is only about 1/30 of that under spot excitation. This dramatic reduction suggests that a moderately extended excitation region allows the radiation field to propagate over a finite distance within the excited system. Although this field is insufficient to sustain ASE, it may facilitate phase correlation among the emitters and the buildup of macroscopic polarization, thereby mitigating the requirements for SF onset. These findings provide further insight into how optical-field propagation contributes to the buildup of macroscopic cooperative coherence and offer a route toward low-threshold, controllable cooperative radiation sources at room temperature.